\documentclass[12pt]{article}

\usepackage{scicite}

\usepackage{graphicx}
\usepackage{color}
\usepackage{soul} 

\newenvironment{sciabstract}{%
\begin{quote} \bf}
{\end{quote}}

\title{Observation of generalized Kibble-Zurek mechanism across a first-order quantum phase transition in a spinor condensate}

\author
{L.-Y. Qiu\thanks{These authors contributed equally to this work.}, H.-Y. Liang\thanks{These authors contributed equally to this work.}%
, Y.-B. Yang\thanks{These authors contributed equally to this work.}, H.-X. Yang, T. Tian,
Y. Xu \footnote{yongxuphy@mail.tsinghua.edu.cn}, L.-M. Duan\footnote{lmduan@tsinghua.edu.cn}\\
Center for Quantum Information, IIIS,\\
Tsinghua University, Beijing 100084,
PR China}

\date{}

\begin{document}

\sethlcolor{red} 

\baselineskip24pt


\maketitle


\begin{sciabstract}
The Kibble-Zurek mechanism provides a unified theory to describe the universal scaling laws in the dynamics when a system
is driven through a second-order quantum phase transition. However, for first-order quantum phase transitions,
the Kibble-Zurek mechanism is usually not applicable. Here, we experimentally demonstrate and theoretically analyze
a power-law scaling in the dynamics of a spin-1 condensate across a first-order quantum phase transition,
when a system is slowly driven from a polar phase to an antiferromagnetic phase. We show that this power-law
scaling can be described by a generalized Kibble-Zurek mechanism.
Furthermore, by experimentally measuring the spin population, we show the power-law scaling
of the temporal onset of spin excitations with respect to the quench rate, which agrees well with our
numerical simulation results. Our results open the door for further exploring the generalized Kibble-Zurek
mechanism to understand the dynamics across first-order quantum phase transitions.
\end{sciabstract}

\section*{INTRODUCTION}

Nonequilibrium dynamics across phase transitions plays a crucial role in various areas of physics
ranging from cosmology to condensed matter~\cite{polkovnikov2011colloquium}. At zero temperature, the properties of a quantum system are
dictated by its ground state, and the quantum phase transition is driven by quantum fluctuations.
There, at the phase transition point, the energy gap vanishes and the relaxation time diverges,
resulting in the violation of adiabaticity as the system parameter is tuned across the transition point.
The Kibble-Zurek mechanism (KZM) describes the dynamics across the transition point by
three evolution regions: two adiabatic and one impulse regions~\cite{kibble1980some,zurek1985cosmological,zurek1993cosmic,zurek1996cosmological,PhysRevLett.95.035701,zurek2005dynamics,PhysRevB.72.161201}. Specifically, when a system
is far away from the transition point, the relaxation time is sufficiently short so that the system can respond
to the change of a parameter and the dynamics is adiabatic. When the system is tuned to be near the
point, it enters into an impulse region, where the relaxation time is sufficiently long so that the system
cannot adapt to the change and thus remains frozen. After the impulse region, the energy gap becomes large and the
system reenters into an adiabatic region. Based on the KZM, universal scaling laws are predicted
across continuous quantum phase transitions for various quantities, such as topological defects and spin excitations.
The KZM in quantum phase transitions has been experimentally observed in several
systems~\cite{chen2011quantum,braun2015emergence,PhysRevLett.116.155301,clark2016universal,zhang2017defect,PhysRevLett.118.220401,PhysRevA.95.053638,Nature2019Lukin}, such as Bose-Einstein condensates and a programmable Rydberg simulator.

Different from the second-order quantum phase transition, multiple phases coexist at the transition point
for the first-order one. Interestingly, similar to the former, numerical simulations have suggested that scaling laws may also exist in the
dynamics of several first-order phase transitions~\cite{panagopoulos2015off,Zhong2017PRE,coulamy2017dynamics,pelissetto2017dynamic,PhysRevA.95.053638,Shimizu_2018}. However, while the KZM is very successful in the former,
some direct application of the KZM to the first-order transition cannot give a satisfied description of the
scaling law compared to the numerical simulation results,
such as in an extended Bose-Hubbard model~\cite{Shimizu_2018}.
In addition, there has been no experimental evidence for the existence of the
scaling law at the first-order quantum phase transition.

A spinor Bose-Einstein condensate (BEC) provides a versatile platform to study the
nonequilibrium physics, such as spin domains~\cite{PhysRevA.79.043631,PhysRevA.81.053612,PhysRevLett.105.090402,PhysRevA.84.063625,parker2013direct}, topological defects~\cite{sadler2006spontaneous,PhysRevLett.103.250401,PhysRevLett.108.035301,choi2012imprinting},
the KZM through the second-order phase transition~\cite{PhysRevLett.116.155301} and
dynamical quantum phase transitions~\cite{Yang2019PRA,Tian2020PRL}. The condensate is described by a vector order
parameter. Under single-mode approximation, all spin states share the same spatial wave
function so that the spin and spatial degrees
of freedom are decoupled~\cite{14,15}. For an
antiferromagnetic (AFM) sodium condensate, its spin degrees of freedom exhibit a first-order quantum
phase transition between an AFM phase with two $m_F=\pm 1$ levels equally populated and a
polar phase with only the $m_F=0$ level populated ($m_F$ is the magnetic quantum number).
This system therefore provides an ideal platform
to explore the dynamics across the first-order quantum phase transition. Indeed, many interesting
phenomena, such as coarsening dynamics of the instability~\cite{Raman2011PRL},
nematic
and magnetic spin density waves~\cite{Raman2013PRL}, and
dynamical phase transitions~\cite{Yang2019PRA}, have been experimentally observed in
the spinor condensate.

In this paper, we theoretically and experimentally study the scaling law as a quadratic
Zeeman energy is slowly varied from positive to negative values (or from negative
to positive values) across the first-order quantum phase transition between the polar phase
and the AFM phase. Our numerical simulation shows the existence of a power-law scaling
of the temporal onset of the spin excitations with respect to the quench rate.
Similar to the KZM at the continuous quantum phase transition, we find that the dynamics
exhibits two adiabatic and one frozen evolution region, suggesting the existence of the KZM.
For the KZM, the power-law scaling
exponent is directly related to the scaling of the energy gap.
For the conventional one, the
scaling exponent is determined by the energy gap between the ground state and the first
excited state. However, we find that this does not agree with our simulation result.
We therefore generalize the KZM by considering the energy gap between the maximally occupied state (corresponding to
the metastable phase) and its
corresponding first excited state. Using this gap, we find that the predicted
exponent agrees very well with our simulation result.

We further perform experiments in the sodium condensate to show the power-law
scaling of the temporal onset of spin excitations with respect to the quench rate
by measuring the spin population. The experimental results agree well with
our numerical simulations and the generalized KZM. Our result shows the first experimental evidence
for the existence of the power-law scaling in the dynamics across the
first-order quantum phase transition.

\begin{figure}[t]
    \includegraphics[width=\textwidth]{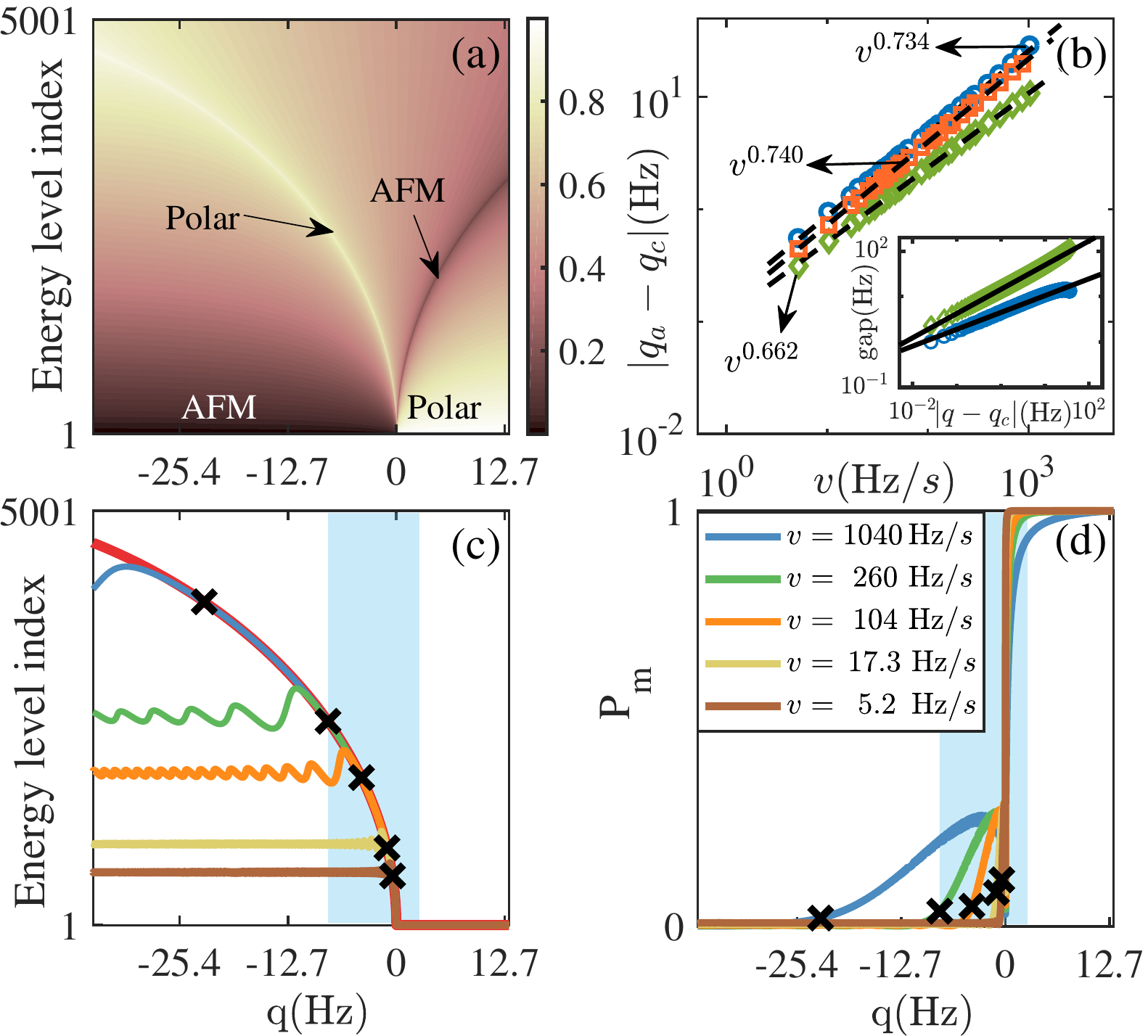}
    \caption{(Color online) (a) $\langle \rho_0\rangle$ for each energy level as a function of $q$.
    The existence of an AFM metastable state for $q>0$ and a polar metastable state for $q<0$ is observed.
    (b) The scaling of the $q$ onset of the spin excitations $q_a$ with respect to the quench rate $v$.
    Orange squares, green diamonds and blue circles are obtained by the numerical simulation, the KZM
    and the generalized KZM, respectively. The inset displays the scaling for the two gaps used in the KZM (green diamonds)
    and the generalized KZM (blue circles) with power-law fitting exponents of $\nu=0.521$ and {\color{red}$\nu=0.371$},
    respectively.
    (c) The evolution of the maximally occupied level $n_{\textrm{max}}(t)$ for distinct $v$ when $q$ is varied from positive to negative values.
    The solid red line depicts the maximally occupied energy level $n_{\textrm{smax}}$ for the initial state. This line coincides with the metastable polar phase as shown in (a).
    (d) The evolution of the probability on the maximally occupied level, i.e., $P_m=|\langle \psi_{n_{\textrm{max}}}(q)|\psi(t)\rangle|^2$,
    for distinct $v$. In (c) and (d), the diagonal crosses label the position $q_a$ where the spin excitations begin
    appearing, calculated by the numerical simulation. In (c) and (d), the filled light blue region shows the frozen region
    for $v=260\,\textrm{Hz}/s$, where the evolving state remains unchanged. We take $c_2=25.4\, \textrm{Hz}$ and the total atom number $N=1.0\times10^{4}$ in the numerical simulation with the energy level index of the Hamiltonian varying from $1,2, \cdots, 5001$.
  }
  \label{fig1}
\end{figure}

\section*{RESULTS}
\subsection*{Theoretical analysis}
We start by considering a spinor BEC, which is well described by the following Hamiltonian under single-mode approximation
\begin{equation}
\hat{H}(q)=c_2 \frac{\hat{L}^2}{2N}+\sum_{m_F=-1}^1(qm_{F}^2-pm_F)\hat{a}_{m_{F}}^{\dagger}\hat{a}_{m_F},
\end{equation}
where $\hat{a}_{m_F}$ ($\hat{a}_{m_F}^\dagger$) is the annihilation (creation) operator for the spin $m_F$ component
corresponding to the hyperfine level $|F=1,m_F\rangle$, $L_\mu=\sum_{m,n}\hat{a}^\dagger_m(f_\mu)_{mn}\hat{a}_n$ is the
condensate's total spin operator along $\mu$ ($\mu=x,y,z$) with $f_\mu$ being the corresponding spin-1 angular momentum matrix,
$c_2$ is the spin-dependent interaction ($c_2>0$ for the antiferromagnetic sodium atoms), $N$ is the total atom number and
$q$ ($p$) is the quadratic (linear) Zeeman energy.

In the absence of the linear Zeeman energy ($p=0$), there are two phases for the ground state: a polar phase with
atoms all occupying the $m_F=0$ level for $q>0$ and the AFM phase with atoms equally occupying the $m_F=\pm 1$ levels for $q<0$~\cite{15}.
If we take mean value $\langle \rho_0\rangle$ with $\rho_0=\hat{a}_0^\dagger\hat{a}_0/N$ as an order parameter, we can clearly see
that $\langle \rho_0\rangle$ abruptly drops from one to zero at $q=0$, showing the
first-order quantum phase transition there [see Fig.~\ref{fig1}(a)].
At the transition point $q_c=0\,\textrm{Hz}$, these two phases coexist.
In fact, near this point, we can observe the existence of the polar phase for $q<0$ and AFM phase for $q>0$ as metastable states,
which is the characteristic of the first-order phase transitions.
In real experiments, $p$ is nonzero. However,
since the Hamiltonian commutes with the total magnetization $\hat{L}_z$, i.e., $[\hat{H}(q),\hat{L}_z]=0$,
the quench dynamics is restricted in the subspace with zero magnetization if we prepare the initial state in the
polar phase and the linear Zeeman term therefore becomes irrelevant.

\begin{figure}[t]
  \includegraphics[width=\textwidth]{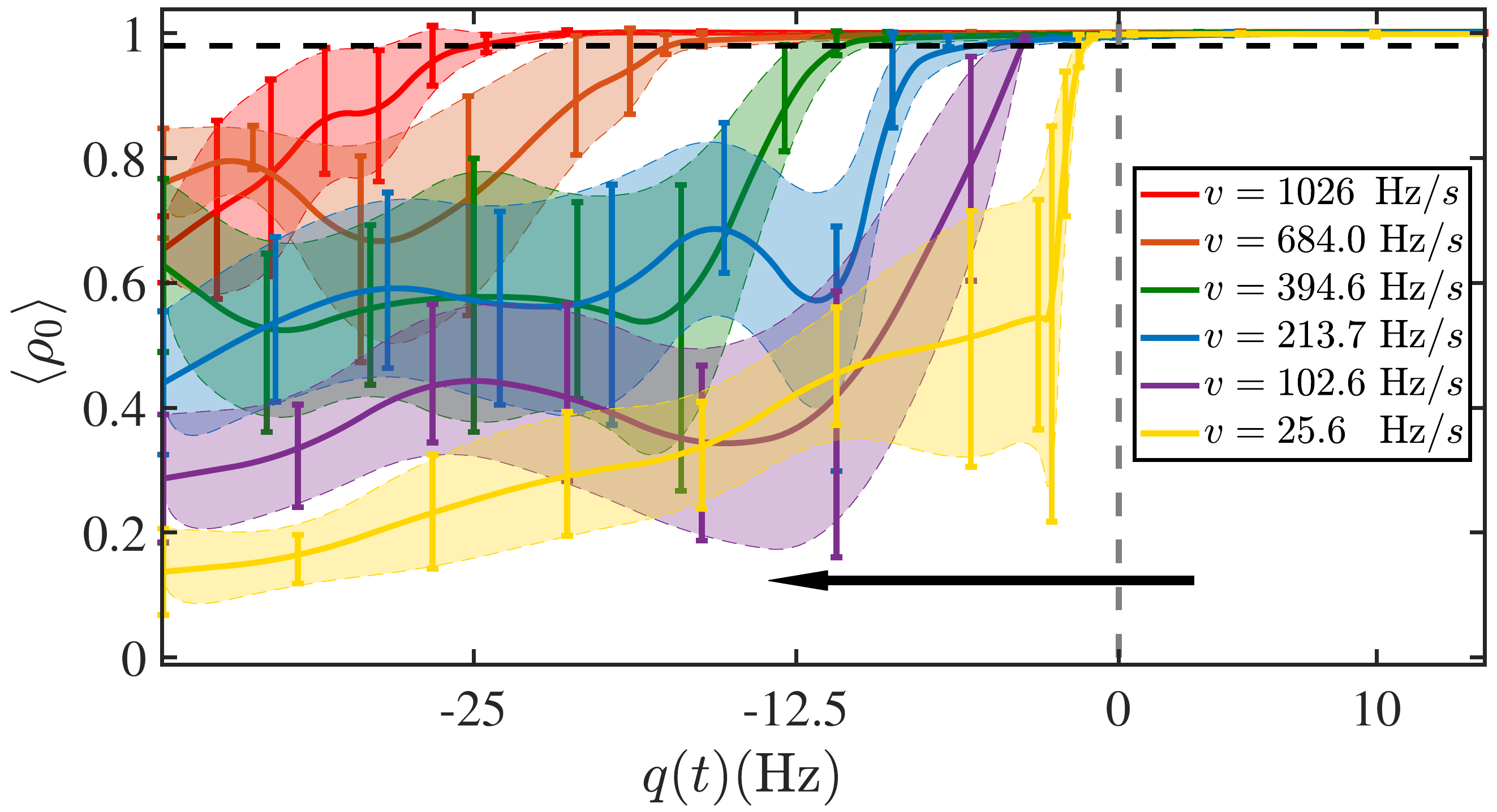}
    \caption{(Color online) Experimentally measured mean value and standard deviation (denoted by the error bar) of $\rho_0$, i.e., $\langle \rho_0\rangle$ and $\Delta \rho_0$ with respect to $q(t)$ as $q(t)$ is slowly varied
    from positive to negative values for a number of ramp rates $v$ with each point repeating 10 times. The horizontal and vertical dashed lines show the
    $\langle \rho_0\rangle$ threshold $\rho_{0c}=0.98$ and the phase transition point $q_c$, respectively. $\langle \rho_0\rangle$ remains unchanged
    in the frozen region until at $q_a$ when it begins to change, entering into the adiabatic region. Here, $c_2=25.5 \pm 1.5\,\textrm{Hz}$.
  }
  \label{fig2} 
\end{figure}

To simulate the scaling in the dynamics across the first-order quantum phase transition, we
start with the ground state of a spinor condensate in the polar phase for positive $q_i$ and
then linearly vary the quadratic Zeeman energy $q$ by $q(t)=q_i-vt$ with $q_i>q_c$, $q_f<q_c$
and $v=(q_i-q_f)/\tau_q$ characterizing the quench rate with $\tau_q$ being the total time as
$q$ changes from $q_i$ to $q_f$.
To numerically simulate the dynamics, we solve the Schr\"odinger equation $i \hbar \, \partial | \psi (t)\rangle /\partial t=\hat{H}(t) |\psi(t)\rangle $ by directly diagonalizing the many-body Hamiltonian with Fock state basis $|N_{+1},N_0,N_{-1}\rangle=\{|N/2,0,N/2\rangle, |N/2-1,2,N/2-1\rangle, \cdots |0,N,0\rangle\}$
(we will take $h=1$ for simplicity hereafter). The time evolution of $\rho_0$ can be obtained by
$\langle\rho_0\rangle(t)={\langle \psi(t)|\rho_0|\psi(t) \rangle}$ for distinct $v$.
In the dynamics across the transition point, spin excitations from the polar state emerge,
which can be reflected by the decrease of $\langle \rho_0 \rangle(t)$ from one. Let $t_a$ be the temporal onset of the spin excitations
and $q_a=q(t=t_a)$ be the critical quadratic Zeeman energy at which the $\langle \rho_0 \rangle(t)$ begins to change.
In Fig.~\ref{fig1}(b), we show the presence of a power-law scaling for $q_a$ with respect to the quench rate $v$ (see the orange squares).

\begin{figure*}[t]
\includegraphics[width=\textwidth]{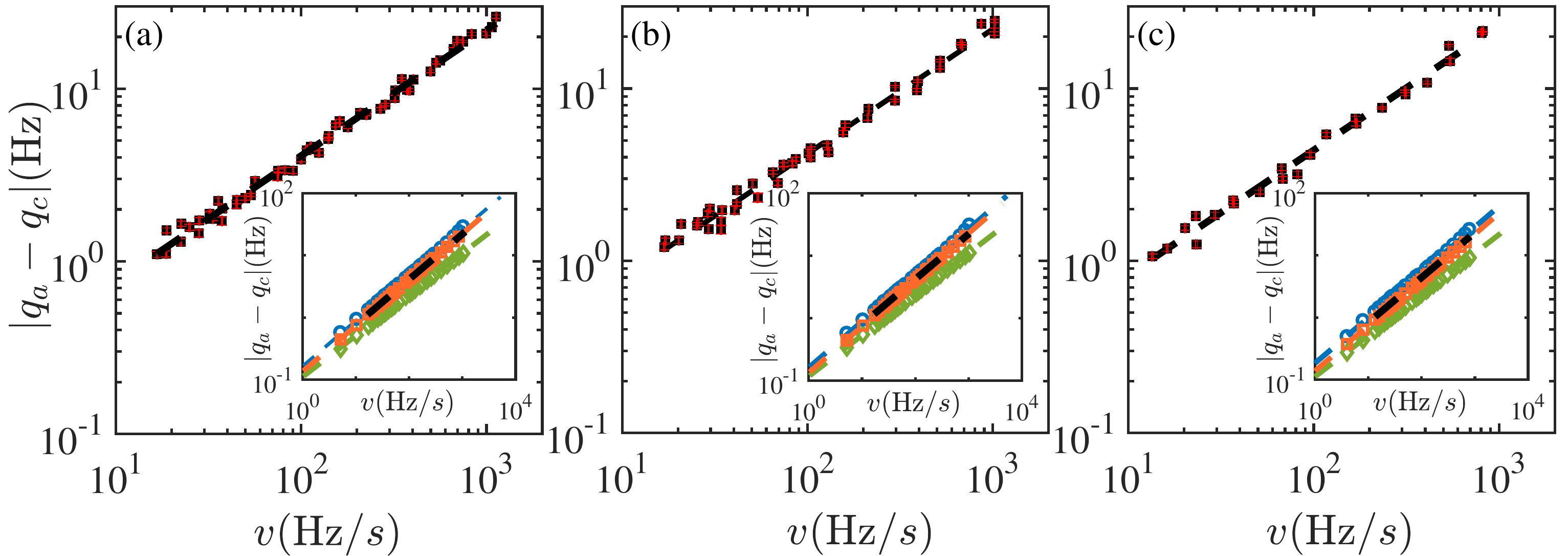}
  \caption{(Color online) Experimentally observed scaling for $|q_a-q_c|$
  with respect to the quench rate $v$ shown
  in the logarithmic scale. In (a-b), $q$ is tuned from around $15\,\textrm{Hz}$ to around $-38\,\textrm{Hz}$ and in (c),
  from around $-12 \,\textrm{Hz}$ to $28 \textrm{Hz}$.
  In (a-c), $c_2=25.5\pm 1.5 \,\textrm{Hz}$, $c_2=23.5\pm 0.7 \,\textrm{Hz}$
  and $c_2=25.2 \pm 0.9\,\textrm{Hz}$, respectively.
  The fitting of the experimental data shows the power-law scaling with the exponent of $0.728 \pm 0.20$ in (a),
  $0.723 \pm 0.25$ in (b) and $0.724 \pm 0.32$ in (c) with $95\%$ confidence boundary. In the insets, we also plot the results of the numerical simulation (orange line),
  the KZM (green line) and
  the generalized KZM (blue line).
  For the numerical simulation, we take {\color{red}$N=1.16\times 10^4$} in (a), {\color{red}$N=0.99\times 10^4$} in (b), and {\color{red}$N=1.07\times 10^4$} in (c) associated with the
  corresponding $c_2$.
  The experimentally observed exponents agree well with the exponents of {\color{red}$0.739$} in (a), {\color{red}$0.744$} in (b) and {\color{red}$0.734$} in (c),
  which are obtained by the numerical calculation. The corresponding exponents predicted by the (generalized) KZM are
  {\color{red}0.662 ($0.733$)}, {\color{red}0.657 ($0.740$)} and {\color{red}0.662 ($0.730$)}, respectively.
  The error of $|q_a-q_c|$ arises from the onset time errors in experiments. For instance, if $\rho_0(t_1)>0.98$ and $\rho_0(t_2)<0.98$, we take $t_a=(t_1+t_2)/2$ with the error of $t_1-t_2$, leading to the error of $q_a$
  being $v(t_1-t_2)$. In experiments, the error is smaller than $0.5\,\textrm{Hz}$ and if $v<52\,\textrm{Hz}/s$, the error is smaller than $0.2\,\textrm{Hz}$.}
\label{fig3} 
\end{figure*}

{\color{red}To delve into the reason underlying the presence of the scaling, let us show the presence of impulse and adiabatic evolution
regions. It is well known that a metastable phase exists across a first-order quantum phase transition as shown in Fig.~\ref{fig1}(a).
Intuitively, when we vary the system parameter $q$ across zero, the evolving state should stay around this metastable state
if the energy gap relative to this metastable state is sufficiently small, suggesting the presence of an impulse region.
Yet, when the energy gap becomes sufficiently large, the state cannot jump to the metastable state of the following $q$ so that $\rho_o$ begins to
decrease, suggesting an entrance into an adiabatic region.
Specifically, in the impulse region, an evolving state remains frozen in the
initial state as time progresses.
In other words,
if the state remains in the initial state, its maximally occupied level
for the evolving state is the same as the maximally occupied level for the initial state.
Here the maximally occupied energy level of
the evolving state is defined as
the $n_{\textrm{max}}(t)$th eigenstate $|\psi_{n_{\textrm{max}}}(q)\rangle$ satisfying
$|\langle \psi_{n_{\textrm{max}}}(q)|\psi(t)\rangle| \ge |\langle \psi_n(q)|\psi(t)\rangle|$ for all $n$
with $|\psi_n(q)\rangle$ being the eigenstate of $\hat{H}(q(t))$, and
the maximally occupied energy level for the initial state is defined as the $n_{\textrm{smax}}$th energy level that has the maximal overlap with the initial state, i.e.,
$|\langle \psi_{n_{\textrm{smax}}}(q)|\psi_0\rangle| \ge |\langle \psi_n(q)|\psi_0\rangle|$ for all $n$.
The latter level coincides with the metastable polar phase with respect to $q$ [see Fig.~\ref{fig1}(a) and (c)], which is consistent with the first-order quantum phase transition.

Interestingly, we find that
when $q$ is varied across zero, the former maximally occupied level $n_{\textrm{max}}(t)$
rapidly increases by following the latter maximally occupied one $n_{\textrm{smax}}$
as shown in Fig.~\ref{fig1}(c),
suggesting the existence of an impulse region where the state remains
frozen.
In contrast, when the system leaves this region, the maximally
occupied level $n_{\textrm{max}}(t)$ begins approaching a fixed level, suggesting the presence of an adiabatic evolution.
For instance, when
$v=260\,\textrm{Hz}/s$, the maximally
occupied level $n_{\textrm{max}}(t)$ follows $n_{\textrm{smax}}$ inside the blue region and then
converges to around the 2510th level in the long time limit [see the green line in Fig.~\ref{fig1}(c)].}

To further demonstrate the existence of impulse and adiabatic regions in the dynamics,
we compute the evolution of the probability of atoms occupying the maximally occupied level, i.e.,
$P_m(q)=|\langle \psi_{n_{\textrm{max}}}(q)|\psi(t)\rangle|^2$.
As shown in Fig.~\ref{fig1}(d), we find that the probability changes rapidly
near the transition point, consistent with the prediction of an impulse evolution, and
remains almost constant in other regions, consistent with the prediction of an adiabatic evolution.
In addition, we mark out $q_a$ as diagonal crosses determined by the numerical simulation,
which agrees well with the $q$ where the system leaves the impulse region and enters the adiabatic
region [see Fig.~\ref{fig1}(c-d)].

The presence of the impulse and adiabatic regions suggests that the scaling law may be accounted for by the KZM.
Suppose that at $t=0$, $q=q_c=0$ and the system is in the polar phase. The $q$ is then linearly varied by $q=-vt$.
Based on the KZM, the critical time when the system begins to respond is determined by $\tau(t_a)=t_a$,
where $\tau(t_a)$ is the relaxation time proportional to $1/\Delta E(t)$ with $\Delta E(t)$ being the energy gap
near the transition point.
We can also determine the critical time $t_a$ by $1/|\Delta E(t)| = |\Delta E(t) / (d\Delta E(t)/dt)|$, after which
the adiabaticity is restored. If the energy gap $\Delta E\propto |q-q_c|^{\nu}$ with $\nu$ being
a positive real number, then the critical time is given by $t_a \propto v^{-\nu/(\nu+1)}$ yielding
$q_a \propto v^{1/(\nu+1)}$.
This shows a power-law scaling of $q_a$ with respect to $v$ and the scaling exponent is determined by
the energy gap.
At the second-order phase transition, the relevant energy gap is the gap between the ground state
and the first excited state labelled as $\Delta E_{12}$. In our system, this energy gap $\Delta E_{12}\propto q^{1/2}$
contributed by the Bogoliubov spin excitations as $q\rightarrow 0$~\cite{15}. This gives us $q_a \propto v^{2/3}$, consistent
with our numerical result $q_a \propto v^{0.662}$ [see Fig.~\ref{fig1}(b)]. It also tells us that the
finite-size effects are very small when $N=1\times 10^4$ (see the Supplementary Materials for details about finite-size effects). However, at the first-order transition point,
the numerical evolution gives us the exponent of $0.740$, which is larger than the value predicted by
the KZM by more than $10\%$. In stark contrast, if the energy gap is taken as
the gap (dubbed the generalized energy gap) between the maximally occupied energy level, i.e., the $n_{\textrm{smax}}$th level and the corresponding first excited state relative to it, i.e., the $(n_{\textrm{smax}}+1)$th level,
we find the exponent of {\color{red}$0.734$}, which agrees well with our numerical result.
This is
due to the different energy gap scaling as shown in the inset of Fig.~\ref{fig1}(b)
{\color{red}[the scaling exponent for the generalized energy gap is $\nu=0.371$].
For the maximally occupied energy level, while there are two gaps relative to it: one with
the next level and the other with the previous level, only the former is relevant since it determines the impulse and
adiabatic regions (see Materials and Methods for details).}
We call this method the generalized KZM.
Yet,
when we apply the generalized KZM to the second-order quantum phase transition,
we find that the result is not as good as the one predicted by the first one,
suggesting the difference between the first-order and second-order quantum
phase transitions (see the Supplementary Materials for details about the KZM across the second-order quantum phase transition).
{\color{red}While the energy gap for a second-order quantum phase transition generically exhibits
a power-law scaling near the critical point, whether the power-law scaling of the generalized energy gap for a
first-order quantum phase transition is universal is still an open question.}

\subsection*{Experimental results}
In experiments, we prepare a sodium BEC in the $3^2S_{1/2}$ $|F=1\rangle$ hyperfine state by evaporation of atoms in an all-optical trap~\cite{Yang2019PRA} and then apply a strong magnetic field gradient to kick the atoms on the $m_F= \pm 1$ levels out of the optical trap, leaving all atoms on the $m_F=0$ level. After that, we hold the BEC atoms in a uniform magnetic field for $3\,\textrm{s}$ to obtain a polar phase under the quadratic Zeeman energy of $q_B=42\,\textrm{Hz}$ induced by the magnetic field.
At the end of the holding, we turn on the microwave pulse with the frequency of $1.7701264\,\textrm{GHz}$ (with the detuning of $-1500\,\textrm{kHz}$ from $|F=1,m_F=0\rangle$ to $|F=2,m_F=0\rangle$) to change the quadratic Zeeman energy to $q_i \simeq 15\,\textrm{Hz}$ (this time is defined as $t=0$). Subsequently, we linearly vary the quadratic Zeeman energy from
$q_i \simeq 15\,\textrm{Hz}$ to $q_f \simeq -38\,\textrm{Hz}$ by ramping up
the amplitude of the microwave field.
During the entire ramping time, we control the microwave power by a
PID system according to the calibration of the quadratic Zeeman energy (see Materials and Methods for details about the
$q$ calibration). As time progresses, we apply the Stern-Gerlach fluorescence imaging to measure $\rho_0 (t)$.
At each time $t$, we repeat $15-20$ measurements to obtain the average $\langle \rho_0\rangle$ over the ensemble.
Fig.~\ref{fig2} displays the observed $\langle \rho_0 \rangle$ as time evolves for a number of ramping rates $v$.
The $q_a$ is taken as the value when $\langle \rho_0\rangle$ drops below the threshold $\rho_{0c}=0.98$.
Evidently, $q_a$ approaches zero as $v$ is decreased.

To experimentally measure the power-law scaling, $q_c$ should be precisely probed. We here employ the quench dynamics to
identify the error of transitions point~\cite{Yang2019PRA} in our calibration.
Besides, we also employ the result to evaluate the error of the $v$ (see Materials and Methods for details about the $q$ error evaluation).

In Fig.~\ref{fig3}(a-b), we plot the observed $|q_a-q_c|$ with respect to $v$ in the logarithmic scale,
showing the existence of a power-law scaling, i.e., $|q_a-q_c|\propto v^{\beta}$.
The fitting of the experimental data gives the exponent of $\beta= 0.728 \pm 0.20$ when $c_2=25.5\pm 1.5\,\textrm{Hz}$ and $\beta= 0.723 \pm 0.25$ when $c_2=23.5\pm 0.7\,\textrm{Hz}$,
 which are slightly different for different $c_2$
due to the finite-size effect.
The experimental results are also in good agreement with the numerical simulation results:
{\color{red}$\beta=0.739$} for the former and {\color{red}$\beta=0.744$} for the latter (see the insets in Fig.~\ref{fig3}).
We also calculate the scaling law determined by the KZM and find that the exponents predicted by the generalized KZM
are {\color{red}$0.733$} for (a) and {\color{red}$0.740$} for (b),
which are closer to the simulation and the experimental results than the exponents of $0.662$
and $0.657$ predicted by the KZM.
In addition, we find that the scaling is not sensitive to the atom loss as we can still observe it
in the presence of $18\%$ atom loss (see Materials and Methods for details about atom loss).

The scaling can also be observed when the system is driven from the AFM phase to the polar phase.
In experiments, we prepare the initial state of the spinor BEC in a nearly AFM state by shining a
$\pi/2$-pulse radio frequency radiation to the BEC on the
$|m_F=0\rangle$ level. We then shine a resonant microwave pulse with
the frequency of $1.7716264\,\textrm{GHz}$ on the atoms for $300\,\textrm{ms}$ to remove the remaining atoms on the $|m_F=0\rangle$ level to obtain an AFM state.
After that, we suddenly switch off this microwave pulse and switch on another one
with the frequency of $1.7701264\,\textrm{GHz}$. By controlling the amplitude of a microwave field,
we are able to linearly vary the quadratic Zeeman energy from around $-12\,\textrm{Hz}$ to around $28\,\textrm{Hz}$.
In Fig.~\ref{fig3}(c), we show the experimentally
measured relation between $|q_a-q_c|$ and $v$, illustrating a power-law scaling with the exponent of
$0.724 \pm 0.32$, which agrees very well with the numerical simulation result of {\color{red}$0.734$}
and the result of {\color{red}$0.730$} predicted by the generalized KZM.

\section*{DISCUSSION}

In summary, we have theoretically and experimentally studied the dynamics across the first-order quantum phase transition
in a spin-1 condensate. We find the existence of a power-law scaling of the temporal onset of the spin excitations with
respect to the quench rate. The scaling is well explained by the generalized KZM.
We further perform an experiment to observe the power-law scaling by measuring the spin populations,
which agrees well with the numerical simulation and the generalized KZM results. Our experiment is the first one to observe the scaling in the dynamics
across the first-order quantum phase transition and hence opens an avenue for
further studying universal scaling
laws for first-order quantum phase transitions both theoretically and experimentally.

\section*{MATERIALS AND METHODS}

\subsection*{The relevant energy gap}
{\color{red}For the maximally occupied energy level (the $n_{\textrm{smax}}$th level), there are two energy gaps relative to it: one gap (labelled as $\Delta E_+$) between this level
and the next level [the ($n_{\textrm{smax}}+1$)th level] and the other (labelled as $\Delta E_-$) between this level 
and the previous level [the ($n_{\textrm{smax}}-1$)th level]. To show that $\Delta E_-$ is not relevant, let us
suppose that $\Delta E_-$ were relevant. 
Let us further suppose that the evolving state occupies the maximally occupied energy level
when we change the $q$ to $q_1<0$. At this $q$, if $\Delta E_-$ is very small compared to the quench rate and $\Delta E_+$ is very large compared to it, then the system should be in the impulse evolution region so that $\rho_0$ should remain unchanged. However, since the evolution is adiabatic with respect to the next level due to the large $\Delta E_+$,
the state cannot evolve to this level when we slightly decrease the $q$, indicating that it cannot evolve to the maximally
occupied energy level given that the level index of the maximally occupied level rises as the $q$ is decreased.
This leads to the decrease of $\rho_0$ as we decrease the $q$, which contradicts with the result that $\rho_0$ should remain unchanged. This conflict shows that the relevant gap is not $\Delta E_-$.}

\subsection*{Calibration of the spin-dependent interaction $c_2$ }
The calibration of the spin-dependent interaction parameter $c_2$ in our experiments is achieved by
applying a widely used spin oscillation procedure as detailed in the following.

In experiments, we first prepare the BEC in the polar state with all atoms occupying the $m_F=0$ level and then apply a radio
frequency radiation to create a coherent state with $\langle\rho_0\rangle=1/2$ and $\langle\rho_{\pm 1}\rangle=1/4$
under a magnetic field, which contributes a quadratic Zeeman energy of $q_B$. After that, the radio frequency radiation
is switched off and the time evolution of the spinor condensates exhibits oscillations~\cite{black2007spinor}. Since the period and amplitude
of the oscillations are determined by $c_2$ and $q_B$, we can obtain $c_2$
by comparing the experimental results with the theoretical ones under a certain $q_B$ with $q_B=q_z B^2$ where $q_z=277\textrm{Hz}/G^2$.

Specifically, we measure the spin oscillation diagram under six different magnetic quadratic Zeeman energy $q_B$, and
evaluate the mean value and standard deviation of $c_2$, i.e., $\overline{c_2}$ and $\delta c_2$ with a narrow range of fluorescence counting number between a low limit $N_L$ and a high limit $N_H$ as shown by the horizontal error bar in Fig. S1.
%


\subsection*{Calibration of the quadratic Zeeman energy $q$}
To calibrate the quadratic Zeeman energy $q$, we measure the Rabi frequency of $\sigma +,\pi,\sigma-$ transitions under a PID microwave power control system in experiments. We also apply the quench dynamics method to evaluate the error in the calibration process of $q$.

In our experiments, the quadratic Zeeman energy is given by $q=q_B+q_M$, where $q_B$ and $q_M$ are generated by
the magnetic field and microwave pulse, respectively. The magnetic Zeeman energy $q_B$ is about $42\,\textrm{Hz}$ in the whole ramping period.
The microwave Zeeman energy is given by
\begin{equation}
q_M=\frac{\delta E_{m_F=+1}+\delta E_{m_F=-1}-2\delta E_{m_F=0}}{2},
\end{equation}
where
\begin{eqnarray}
\delta E_{m_F}&=&\frac{1}{4}\sum_{k=-1,0,+1} \frac{\Omega_{m_F \rightarrow m_F+k}^2}{\delta_{m_F \rightarrow m_F+k}} \\
&=&\frac{1}{4}\sum_{k=-1,0,+1} \frac{\Omega_{m_F \rightarrow m_F+k}^2}{\delta_0-[(m_F+k)g_F-(- m_F)g_F]\mu_B B}
\end{eqnarray}
with $\Omega_{m_F \rightarrow m_F+k}$ being the resonant Rabi frequency for the transition from $|F=1,m_F\rangle$
to $|F=2,m_F+k\rangle$ and $\delta_0$ being the microwave detuning for the transition from
$|F=1,m_F=0\rangle$ to $|F=2,m_F=0\rangle$.


In experiments, we measure three Rabi frequencies of $\sigma_+$, $\pi$ and $\sigma_-$ transitions
corresponding to $\Omega_{m_F=0\rightarrow m_F=1}$, $\Omega_{m_F=-1\rightarrow m_F=-1}$
and $\Omega_{m_F=0\rightarrow m_F=-1}$, respectively, and then determine the $q_M$ based on the above formula.
The detuning of the microwave pulse $\delta_0$ is precisely controlled by the Keysight E8663D PSG RF Analog Signal Generator.
Without a PID system, its power requires more than $1$ s to reach a stable value (after the RF amplifier ZHL-30W-252-S+),
causing an error of $q$ about $-3\,\textrm{Hz}$. We therefore apply a PID system to
shorten the time for the microwave power to reach a set value $V_{\textrm{set}}$ to less than $100\mu s$.
The Rabi frequencies are measured during $130\mu s-300\mu s$ after the microwave pulse is switched on.

In Fig. S2, we plot the result of $q$ based on the experimentally measured
Rabi frequencies at six distinct $V_{\textrm{set}}$ with frequency detuning $\delta_0=-1500\,\textrm{kHz}$.
The figure also shows the fitting of these data by a parabola (see the dashed red line) and with this fitting line, the $V_{\textrm{set}}$ is controlled following the line
shown in
Fig. S2
(b) to realize the linear change of the $q$.

In the following, we apply the quench dynamics
to measure the quantum phase transition point and
evaluate the $q$ calibration error.
We first prepare the BECs in the polar phase under a positive $q_i$ and then suddenly quench the $q$ to
$q_f$. If $q_f$ is positive, the atoms remain on the $m_F=0$ level after $500\,\textrm{ms}$ evolution, and if $q_f$ is negative,
the atoms on the $m_F=\pm1$ levels show up after $500\,\textrm{ms}$ evolution.
In experiments, $\langle \rho_0\rangle$ is measured after this period of time
for distinct $V_{\textrm{set}}$ as the microwave frequency is suddenly tuned to $\nu_f$.
To find the transition point,
we control the $\nu_f$ to find the minimum $\nu_1$ so that
$\langle \rho_{0}\rangle$ remains unchanged and
the maximum $\nu_2$ so that $\langle \rho_{0}\rangle$ is decreased.
Note that the $q$ decreases as the frequency $\nu_f$ is increased with $\delta_0$ varying from $-2000\,\textrm{kHz}$ to $-1300\,\textrm{kHz}$.
For these two frequencies $\nu_{1,2}$, we calculate the quadratic Zeeman energy $q_1$ and $q_2$, respectively,
under the $V_\textrm{set}$.


In table S1, we show the mean value $\overline{q_{1,2}}$ and standard deviation $\delta q_{1,2}$ of $q_{1,2}$
based on the quench dynamics data in one month. The error of the $q$ leads to the error of the $v$ as $\delta v=\sqrt{\delta q_1(V_{\textrm{set}}=500\textrm{mV})^2+\delta q_1(V_{\textrm{set}}=900\textrm{mV})^2}/\tau_q$, where $\delta q_1(V_{\textrm{set}}=500\textrm{mV})$ and $\delta q_1(V_{\textrm{set}}=900\textrm{mV})$ are the standard deviations for $V_{\textrm{set}}=500\textrm{mV}$ and $V_{\textrm{set}}=900\textrm{mV}$, respectively.

\subsection*{The effects of atom loss}
In experiments, atom loss occurs due to the microwave and optical radiation.
In Fig. S3, we display the amount of atom loss for different quench rates, showing that the
amount increases when the $q$ is linearly decreased achieved by controlling the microwave amplitude and it
also increases for smaller $v$. Specifically, when $v=17.1\,\textrm{Hz/s}$, which is the slowest quench rate in the experiments,
the amount of atom loss is roughly $18\%$ at the end of the ramp and $10\%$ near the $q_c=0\,\textrm{Hz}$ point. Despite the presence of
atom loss, it does not have obvious effects on our measured scaling property as shown in Fig. 3 in the main text.

\section*{Acknowledgments}
We thank Yingmei Liu, Ceren Dag, and Anjun Chu for helpful discussions. Funding: This work was supported by the Frontier Science Center for Quantum Information of the Ministry of Education of China, Tsinghua University Initiative Scientific Research Program, and the National key Research and Development Program of China (2016YFA0301902). Y. X. also acknowledges the support
by the start-up fund from Tsinghua University,
the National Thousand-Young-Talents Program and the National Natural Science Foundation
of China (11974201).
Author contributions:
Competing interests: The authors declare that they have no competing interests.
Data and materials availability: All data needed to evaluate the conclusions in the paper are present in the paper and/or the Supplementary Materials. Additional data related to the paper are available from authors upon request.

\section*{Supplementary materials}
Supplementary Text \\
Section S1. Finite-size effects \\
Section S2. The KZM across the second-order quantum phase transition \\
Fig. S1. Calibration of the spin-dependent interaction $c_2$.\\
Fig. S2. The experimental power control of the microwave pulse.\\
Fig. S3. Atom loss results.\\
Fig. S4. Effects of the atom number on the exponent.\\
Fig. S5. The scaling law across a second-order quantum phase transition.\\
Table S1. The error evaluation of $q$.\\


\clearpage


\begin{thebibliography}{10}

\bibitem{polkovnikov2011colloquium}
A. Polkovnikov, K. Sengupta, A. Silva, M. Vengalattore.
\newblock Colloquium: Nonequilibrium dynamics of closed interacting quantum
  systems.
\newblock {\em Rev. Mod. Phys.} \textbf{83}, 863 (2011).

\bibitem{kibble1980some}
T.~W. B. Kibble.
\newblock Some implications of a cosmological phase transition.
\newblock {\em Phys. Rep.} \textbf{67}, 183-199 (1980).

\bibitem{zurek1985cosmological}
W.~H. Zurek.
\newblock Cosmological experiments in superfluid helium?
\newblock {\em Nature} \textbf{317}, 505-508 (1985).

\bibitem{zurek1993cosmic}
W.~H. Zurek.
\newblock Cosmic strings in laboratory superfluids and the topological remnants
  of other phase transitions.
\newblock {\em Acta Phys. Polon.} \textbf{24}, 1301-1311 (1993).

\bibitem{zurek1996cosmological}
W.~H. Zurek.
\newblock Cosmological experiments in condensed matter systems.
\newblock {\em Phys. Rep.} \textbf{276}, 177--221 (1996).

\bibitem{PhysRevLett.95.035701}
B.~Damski.
\newblock The Simplest Quantum Model Supporting the Kibble-Zurek Mechanism of Topological Defect Production: Landau-Zener Transitions from a New Perspective.
\newblock {\em Phys. Rev. Lett.} \textbf{95}, 035701 (2005).

\bibitem{zurek2005dynamics}
W.~H. Zurek, U. Dorner, P. Zoller.
\newblock Dynamics of a Quantum Phase Transition.
\newblock {\em Phys. Rev. Lett.} \textbf{95}, 105701 (2005).

\bibitem{PhysRevB.72.161201}
A. Polkovnikov.
\newblock Universal adiabatic dynamics in the vicinity of a quantum critical
  point.
\newblock {\em Phys. Rev. B} \textbf{72}, 161201 (2005).

\bibitem{chen2011quantum}
D. Chen, M. White, C. Borries, B. DeMarco.
\newblock Quantum Quench of an Atomic Mott Insulator.
\newblock {\em Phys. Rev. Lett.} \textbf{106}, 235304 (2011).

\bibitem{braun2015emergence}
S. Braun, M. Friesdorf, S.~S. Hodgman, M. Schreiber, J.~P.
  Ronzheimer, A. Riera, M. D.~Rey, I. Bloch, J. Eisert,
  U. Schneider.
\newblock Emergence of coherence and the dynamics of quantum phase transitions.
\newblock {\em PNAS} \textbf{112}, 3641-3646 (2015).

\bibitem{PhysRevLett.116.155301}
M.~Anquez, B.~A. Robbins, H.~M Bharath, M.~Boguslawski, T.~M. Hoang, M.~S.
  Chapman.
\newblock Quantum Kibble-Zurek Mechanism in a Spin-1 Bose-Einstein Condensate.
\newblock {\em Phys. Rev. Lett.} \textbf{116}, 155301 (2016).

\bibitem{clark2016universal}
L.~W. Clark, L. Feng, C. Chin.
\newblock Universal space-time scaling symmetry in the dynamics of bosons
  across a quantum phase transition.
\newblock {\em Science} \textbf{354}, 606-610 (2016).

\bibitem{zhang2017defect}
J. Zhang, F.~M. Cucchietti, R. Laflamme, D. Suter.
\newblock Defect production in non-equilibrium phase transitions: experimental investigation of the Kibble-Zurek mechanism in a two-qubit quantum simulator.
\newblock {\em New J. Phys.} \textbf{19}, 043001 (2017).

\bibitem{PhysRevLett.118.220401}
B.~M. Anderson, L.~W. Clark, J. Crawford, A. Glatz, I.~S. Aranson, P. Scherpelz, L. Feng, C. Chin, K.~Levin.
\newblock Direct Lattice Shaking of Bose Condensates: Finite Momentum Superfluids.
\newblock {\em Phys. Rev. Lett.} \textbf{118}, 220401 (2017).

\bibitem{PhysRevA.95.053638}
S. Kang, S.~W. Seo, J.~H. Kim, Y.~Shin.
\newblock Emergence and scaling of spin turbulence in quenched antiferromagnetic spinor Bose-Einstein condensates.
\newblock {\em Phys. Rev. A} \textbf{95}, 053638 (2017).

\bibitem{Nature2019Lukin}
A. Keesling, A. Omran, H. Levine, H. Bernien, H. Pichler,
  S. Choi, R. Samajdar, S. Schwartz, P. Silvi, S. Sachdev,
  P. Zoller, M. Endres, M. Greiner, V. Vuleti\'c, M. D. Lukin.
\newblock Quantum Kibble-Zurek mechanism and critical dynamics on a programmable Rydberg simulator.
\newblock {\em Nature} \textbf{568}, 207-211 (2019).

\bibitem{panagopoulos2015off}
H. Panagopoulos, E. Vicari.
\newblock Off-equilibrium scaling behaviors across first-order transitions.
\newblock {\em Phys. Rev. E} \textbf{92}, 062107 (2015).

\bibitem{Zhong2017PRE}
N. Liang, F. Zhong.
\newblock Renormalization-group theory for cooling first-order phase transitions in Potts models.
\newblock {\em Phys. Rev. E} \textbf{95,} 032124 (2017).

\bibitem{coulamy2017dynamics}
I.~B. Coulamy, A. Saguia, M.~S. Sarandy.
\newblock Dynamics of the quantum search and quench-induced first-order phase transitions.
\newblock {\em Phys. Rev. E} \textbf{95}, 022127 (2017).

\bibitem{pelissetto2017dynamic}
A. Pelissetto, E. Vicari.
\newblock Dynamic Off-Equilibrium Transition in Systems Slowly Driven across Thermal First-Order Phase Transitions.
\newblock {\em Phys. Rev. Lett.} \textbf{118}, 030602 (2017).


\bibitem{Shimizu_2018}
K. Shimizu, T. Hirano, J. Park, Y. Kuno, I. Ichinose.
\newblock Dynamics of first-order quantum phase transitions in extended Bose-Hubbard model: from density wave to superfluid and vice versa.
\newblock {\em New J. Phys.} \textbf{20}, 083006 (2018).

\bibitem{PhysRevA.79.043631}
S.~R. Leslie, J.~Guzman, M.~Vengalattore, J.~D. Sau, M.~L. Cohen,  D.~M. Stamper-Kurn.
\newblock Amplification of Fluctuations in a Spinor Bose Einstein Condensate.
\newblock {\em Phys. Rev. A} \textbf{79}, 043631 (2009).

\bibitem{PhysRevA.81.053612}
M.~Vengalattore, J.~Guzman, S.~R. Leslie, F.~Serwane, D.~M. Stamper-Kurn.
\newblock Periodic spin textures in a degenerate $F=1$ $^{87}\mathrm{Rb}$ spinor Bose gas.
\newblock {\em Phys. Rev. A} \textbf{81}, 053612 (2010).

\bibitem{PhysRevLett.105.090402}
J. Kronj\"ager, C. Becker, P. Soltan-Panahi, K. Bongs, K. Sengstock.
\newblock Spontaneous Pattern Formation in an Antiferromagnetic Quantum Gas.
\newblock {\em Phys. Rev. Lett.} \textbf{105}, 090402 (2010).

\bibitem{PhysRevA.84.063625}
J.~Guzman, G.-B. Jo, A.~N. Wenz, K.~W. Murch, C.~K. Thomas, D.~M.
  Stamper-Kurn.
\newblock Long-time-scale dynamics of spin textures in a degenerate $F=1$
  ${}^{87}$Rb spinor Bose gas.
\newblock {\em Phys. Rev. A} \textbf{84}, 063625 (2011).

\bibitem{parker2013direct}
C.~V. Parker, L. Ha, C. Chin.
\newblock Direct observation of effective ferromagnetic domains of cold atoms
  in a shaken optical lattice.
\newblock {\em Nat. Phys.} \textbf{9}, 769-774 (2013).

\bibitem{sadler2006spontaneous}
L. E.~Sadler, J. M.~Higbie, S. R.~Leslie, M.~Vengalattore, D. M.~Stamper-Kurn.
\newblock Spontaneous symmetry breaking in a quenched ferromagnetic spinor Bose-Einstein condensate.
\newblock {\em Nature}  \textbf{443}, 312-315 (2006).

\bibitem{PhysRevLett.103.250401}
L.~S. Leslie, A.~Hansen, K.~C. Wright, B.~M. Deutsch, N.~P. Bigelow.
\newblock Creation and Detection of Skyrmions in a Bose-Einstein Condensate.
\newblock {\em Phys. Rev. Lett.} \textbf{103}, 250401 (2009).

\bibitem{PhysRevLett.108.035301}
J. Choi, W. J. Kwon, Y. Shin.
\newblock Observation of Topologically Stable 2D Skyrmions in an Antiferromagnetic Spinor Bose-Einstein Condensate.
\newblock {\em Phys. Rev. Lett.} \textbf{108}, 035301 (2012).

\bibitem{choi2012imprinting}
J. Choi, W.~J. Kwon, M. Lee, H. Jeong, K. An,
  Y. Shin.
\newblock Imprinting Skyrmion spin textures in spinor Bose-Einstein condensates.
\newblock {\em New J. Phys.} \textbf{14}, 053013 (2012).

\bibitem{Yang2019PRA}
H.-X. Yang, T.~Tian, Y.-B. Yang, L.-Y. Qiu, H.-Y. Liang, A.-J. Chu, C.~B.
  Da\ifmmode~\breve{g}\else \u{g}\fi{}, Y.~Xu, Y.~Liu, L.-M. Duan.
\newblock Observation of dynamical quantum phase transitions in a spinor
  condensate.
\newblock {\em Phys. Rev. A} \textbf{100}, 013622 (2019).

\bibitem{Tian2020PRL}
T. Tian, H.-X. Yang, L.-Y. Qiu, H.-Y. Liang, Y.-B. Yang, Y. Xu, L.-M. Duan.
\newblock Observation of Dynamical Quantum Phase Transitions with Correspondence in an Excited State Phase Diagram.
\newblock {\em Phys. Rev. Lett.} \textbf{124}, 043001 (2020).

\bibitem{14}
Y. Kawaguchi, M. Ueda.
\newblock Spinor Bose-Einstein condensates.
\newblock {\em Phys. Rep.} \textbf{520}, 253-381 (2012).

\bibitem{15}
D.~M. Stamper-Kurn, M. Ueda.
\newblock Spinor Bose gases: Symmetries, magnetism, and quantum dynamics.
\newblock {\em Rev. Mod. Phys.} \textbf{85}, 1191-1244 (2013).

\bibitem{Raman2011PRL}
E.~M. Bookjans, A.~Vinit, C.~Raman.
\newblock Quantum Phase Transition in an Antiferromagnetic Spinor Bose-Einstein Condensate.
\newblock {\em Phys. Rev. Lett.}  \textbf{107}, 195306 (2011).

\bibitem{Raman2013PRL}
A.~Vinit, E.~M. Bookjans, C.~A.~R. S\'a~de Melo, C.~Raman.
\newblock Antiferromagnetic Spatial Ordering in a Quenched One-Dimensional Spinor Gas.
\newblock {\em Phys. Rev. Lett.} \textbf{110}, 165301 (2013).

\bibitem{black2007spinor}
A.~T. Black, E. Gomez, L.~D. Turner, S. Jung, P.~D. Lett.
\newblock Spinor Dynamics in an Antiferromagnetic Spin-1 Condensate.
\newblock {\em Phys. Rev. Lett.} \textbf{99}, 070403 (2007).

\end{thebibliography}
\end{document}